\documentclass[twocolumn,amsmath,amssymb,prb,notitlepage]{revtex4-1}
\usepackage[T1]{fontenc}
\usepackage[utf8]{inputenc}
\usepackage[english]{babel}
\usepackage{newtxtext,newtxmath}
\usepackage{graphicx}% Include figure files
\usepackage[usenames,dvipsnames]{xcolor} %Needed for comments % Remove for publication
\usepackage[colorlinks=true,linkcolor=blue,citecolor=blue,urlcolor=blue]{hyperref} % For reference links
\usepackage{soul} % Cross out text \sout % Remove for publication
\usepackage{pgfplots}
\pgfplotsset{compat=1.15}
\usetikzlibrary{pgfplots.groupplots}
\usepgfplotslibrary{patchplots}
\usepackage[outline]{contour} % For Plots
\contourlength{1.2pt}   %For Plots
\usetikzlibrary{external}    % For pdf generation of tex images
\bibpunct{[}{]}{,}{n}{}{}

\newcommand{\IM}{\operatorname{Im}}

\newcommand{\Ang}{\,\text{\AA}}
\newcommand{\meV}{\,\mathrm{meV}}
\newcommand{\eV}{\,\mathrm{eV}}

\newcommand{\stress}{\sigma}
\newcommand{\strain}{\epsilon}
\newcommand{\ElasticT}{C}
\newcommand{\GPa}{\,\mathrm{GPa}}
\newcommand{\RH}{R_\mathrm{h}}
\newcommand{\RB}{R_\mathrm{b}}
\newcommand{\Xih}{\Xi_\mathrm{h}}
\newcommand{\Xib}{\Xi_\mathrm{b}}
\newcommand{\Xiu}{\Xi_\mathrm{u}}

\newcommand{\figref}[1]{Fig.\,\ref{#1}}
\newcommand{\equref}[1]{Eq.\,\eqref{#1}}
\newcommand{\secref}[1]{Sec.\,\ref{#1}}
\newcommand{\tabref}[1]{Table\,\ref{#1}}
\newcommand{\citeref}[1]{Ref.\,[\onlinecite{#1}]}

\newcommand{\kv}{\mathbf{k}}
\newcommand{\id}{\text{id}}

\newcommand{\new}[1]{\textcolor{blue}{#1}}
\begin{document}

%Strain engineering of lonsdaleite Ge: an \textit{ab initio} study (Title for PRB)

\title{Efficient strain-induced light emission in lonsdaleite germanium}

\author{Jens Ren\`{e} Suckert}
\author{Claudia R\"{o}dl}
\author{J\"{u}rgen Furthm\"{u}ller}
\author{Friedhelm Bechstedt}
\author{Silvana Botti}
\affiliation{Institut f\"{u}r Festk\"{o}rpertheorie und -optik, Friedrich-Schiller-Universit\"{a}t Jena, Max-Wien-Platz 1, 07743 Jena, Germany}
\affiliation{European Theoretical Spectroscopy Facility}

\date{\today}% It is always \today, today,

\begin{abstract}
Hexagonal germanium in the lonsdaleite structure has a direct band gap, but it is not an efficient light emitter due to the vanishing oscillator strength of electronic transitions at the fundamental gap. Transitions involving the second lowest conduction band are instead at least three orders of magnitude stronger. The inversion of the two lowest conduction bands would therefore make hexagonal germanium ideal for optoelectronic applications.
In this work, we investigate the possibility to achieve this band inversion by applying strain.
%with space group $\mathrm{P}6_3/mmc$ 
To this end we perform \textit{ab initio} calculations of the electronic band structure and optical properties of strained hexagonal germanium, using density functional theory with the modified Becke-Johnson exchange-correlation functional and including spin-orbit interaction.
We consider hydrostatic pressure, uniaxial strain along the hexagonal $c$~axis, as well as biaxial strain in planes perpendicular to and containing the hexagonal $c$~axis to simulate the effect of a substrate.
We find that the conduction-band inversion, and therefore the transition from a pseudo-direct to a direct band gap, is attainable for moderate tensile uniaxial strain parallel to the lonsdaleite $c$~axis.
\end{abstract}

\maketitle
%%%%%%%%%%%%%%%%%%%%%%%%%%%%%%%%%%%%%%%%%%%%%%%%%%%%%%%%%%%%%%%%%%%%
\section{Introduction} \label{sec1}
%%%%%%%%%%%%%%%%%%%%%%%%%%%%%%%%%%%%%%%%%%%%%%%%%%%%%%%%%%%%%%%%%%%%
Integration of silicon-based active optical devices into the complementary metal-oxide semiconductor (CMOS) technology is hampered by the indirect band gap of the diamond-structure phase of silicon (Si), germanium (Ge), and SiGe alloys~\cite{Miller2009,Srinivasa2014}. In fact, none of these semiconductors can emit light efficiently, and therefore they are not suitable for applications that require optoelectronic functionalities. Using heterogeneous integration, III–V semiconductors can be implemented as active light sources onto chips~\cite{Miller2009}. However, it would be desirable to have light sources available that are chemically compatible with Si, tolerated by the CMOS technology, and capable of emitting light efficiently. It is therefore an important technological challenge to achieve low-threshold lasing in group-IV semiconductors.

Strain and alloying are the most investigated strategies to tailor absorption and emission properties of Si and Ge~\cite{Nam2014,Geiger2015}. Another explored way is to search for non-diamond-structure polymorphs with a modified band structure. Indeed, this is a presently very active field of research for both group-IV and III-V semiconductors~\cite{joyce_etal_2011,caroff_bolinsson_johansson_2011,barrigon_etal_2019,galvaotizei_amato_2020}. The hexagonal lonsdaleite phase of Si (hex-Si) with space group $P6_3/mmc$ can be stabilized under ambient conditions using different growth techniques~\new{\cite{zhang_etal_1999,fontcubertaimorral_etal_2007,hauge_etal_2015,vincent_etal_2018}}. Even if the size of the direct gap at $\Gamma$ is reduced to $1.6\eV$ by band folding~\cite{Roedl2015}, the fundamental gap remains indirect with the conduction-band minimum (CBM) at the $M$~point of the hexagonal Brillouin zone (BZ). Biaxial tensile strain larger than $4\,\%$ has been predicted to transform hex-Si into a direct semiconductor~\cite{Roedl2015}. 

Ge also has an excellent CMOS compatibility~\cite{Miller2009}. The band structure of Ge in the cubic diamond structure features an indirect band gap with the CBM at the $L$~point and a direct gap at $\Gamma$ that is only about $0.1\eV$ larger~\cite{Laubscher2015,Roedl2019}. Ge in the hexagonal lonsdaleite crystal structure (hex-Ge) has been predicted to have a direct gap of about $0.3\eV$ at the $\Gamma$~point \cite{Raffy2002,De2014,Kaewmaraya2017,Chen2017,Roedl2019}. Experimentally, hex-Ge has been obtained by strain-induced phase transformation \cite{vincent_etal_2014} and low-pressure ultraviolet laser ablation~\cite{Zhang2000,Haberl2014}, for instance. Alternatively, high-quality crystals of hex-Ge or hex-SiGe alloys can be grown on templates of wurtzite (wz) GaP or GaAs nanowires \cite{Ikaros2017,fadaly2020direct}. Recently, direct-band-gap emission has been demonstrated for Ge-rich hex-SiGe alloys, with an emission wavelength tunable by controlling the alloy composition~\cite{fadaly2020direct}. The possibility to grow bent nanowires, through adjustment of the core–shell sizes, opens the way to the engineering of the optical properties through strain~\cite{Assali2019,Greenberg2019}. In contrast to bulk crystals, it is in this case easier to obtain uniaxial strain and biaxial symmetry-reducing strains on the wire facets. In the following, we will consider strain that can be experimentally realized growing high-quality crystalline core-shell nanowires.

\begin{figure}
    \centering
    %\tikzsetnextfilename{Images_pdf/matrix_elements}
    \includegraphics[width=.48\textwidth]{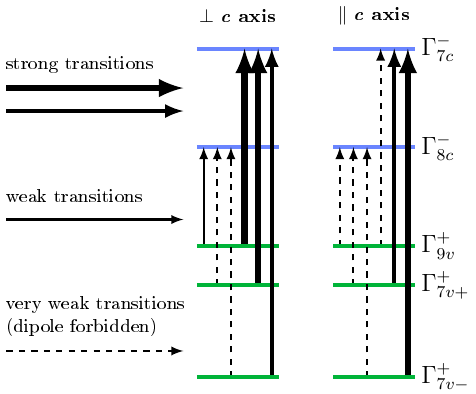}
    \caption{
    Visualization of the optical oscillator strengths between three highest valence bands (green) and the two lowest conduction (blue) of hex-Ge at $\Gamma$. The relative distances between the energy levels are true to scale. The nomenclature of the high-symmetry states follows the double-group notation introduced in Ref.~\cite{Roedl2019}.
    }
    \label{fig:matrix_elements}
\end{figure}

The energy levels of hex-Ge at the $\Gamma$~point are shown in Fig.~\ref{fig:matrix_elements}. The CBM possesses $\Gamma_{8c}^-$ symmetry. In agreement with group theory~\cite{Tronc1999}, optical dipole transitions from the three highest valence bands with symmetries $\Gamma_{9v}^+$, $\Gamma_{7v+}^+$, and $\Gamma_{7v-}^+$ to the lowest conduction band are dipole forbidden except for the transition $\Gamma_{9v}^+\rightarrow\Gamma_{8c}^-$ that is dipole allowed for light polarization perpendicular to the hexagonal $c$~axis~\cite{Roedl2019}. However, its oscillator strength is three orders of magnitude smaller than the oscillator strength of typical dipole-active transitions in semiconductors~\cite{Roedl2019}. Thus, hex-Ge can be called a pseudo-direct semiconductor that, despite having a direct band gap, is not optically active at the band-gap energy. However, there is a second conduction band with $\Gamma_{7c}^-$ symmetry about $0.3\eV$ above the $\Gamma_{8c}^-$ CBM. Except for the $\Gamma_{9v}^+\rightarrow \Gamma_{7c}^-$ transition for light polarization parallel to the $c$~axis, all transitions from the three highest valence bands to $\Gamma_{7c}^-$ are (strongly) dipole active~\cite{Tronc1999,Roedl2019}. 

In this work, we use accurate \textit{ab initio} calculations to investigate if the ordering of the two lowest conduction bands can be inverted by applying moderate lattice strain to obtain a CBM with $\Gamma_{7c}^-$ character and make hex-Ge an efficient light emitter. We also explore the possibility to tune the wavelength of light emission by strain~\cite{Signorello2014}.
A brief review of the used methods is given in \secref{sec2}. The influence of strain on the atomic structure is studied in \secref{sec3}. Section~\ref{sec4} is dedicated to the analysis of strain effects on the electronic band structure. In \secref{sec5}, we focus on the optical properties of strained hex-Ge. Finally, in \secref{sec6}, we present a summary and draw conclusions.

%%%%%%%%%%%%%%%%%%%%%%%%%%%%%%%%%%%%%%%%%%%%%%%%%%%%%%%%%%%%%%%%%%%%%%
\section{Computational details} \label{sec2}
%%%%%%%%%%%%%%%%%%%%%%%%%%%%%%%%%%%%%%%%%%%%%%%%%%%%%%%%%%%%%%%%%%%%%%

All calculations were performed in the framework of density-functional theory (DFT) as implemented in the Vienna \textit{Ab-initio} Simulation Package (\textsc{Vasp}) \cite{Kresse1996_1,Kresse1996_2}. The wave functions are described using the projector-augmented wave method \cite{Kresse1999} with a plane-wave cutoff of $500\eV$. The Ge\,$3d$ electrons are treated as valence electrons and spin-orbit coupling is included in all calculations. Unless otherwise noted, a $12\times12\times6$ $\Gamma$-centered $\kv$-point mesh is used for the BZ integration.

For the calculation of structural and elastic properties under strain, exchange and correlation are described in the generalized gradient approximation (GGA) using the PBEsol~\cite{Perdew2008} functional. It has been previously shown that the PBEsol functional provides excellent lattice parameters for hexagonal Ge~\cite{Roedl2019,fadaly2020direct}. The strained structures were optimized under constraint until the Hellmann-Feynman forces for all atomic coordinates were below $1\meV/\text{\AA}$. Symmetry-reducing strain on hexagonal nanowire facets is simulated in an orthorhombic supercell of the primitive hexagonal cell using a $12\times6\times 6$ $\kv$-point grid.

Electronic band structures were obtained with the meta-GGA functional MBJLDA, i.e.\ the modified Becke-Johnson exchange potential~\cite{Becke2006,Tran2007,Tran2009} together with the local-density approximation (LDA) for correlation. This meta-GGA functional yields excellent band gaps at low computational cost \cite{Laubscher2015,Chen2017,Roedl2019,borlido2019large}, even though deviations from experiment or more sophisticated approaches (e.g.\ hybrid functionals or many-body perturbation theory) occur for electronic states further away from the band gap. Since we are mainly interested in the near-gap electronic structure here, this does not pose a problem for the present study. We carefully validated the choice of this functional for electronic-structure calculations of hex-Ge in Ref.~\cite{Roedl2019}.

Optical spectra and radiative lifetimes were calculated in the independent-particle approximation starting from the meta-GGA electronic band structure using a denser BZ sampling with $60\times60\times30$ $\Gamma$-centered $\kv$~points. 
% SB Checked in Rene's data: the k-point grid is 60x60x30. However the matrix elements are calculated only with a 12x12x6 grid.
% CR Does not matter for the matrix elements as we only take the Gamma one. The matrix elements for the spectra are on the dense grid.

%%%%%%%%%%%%%%%%%%%%%%%%%%%%%%%%%%%%%%%%%%%%%%%%%%%%%%%%%%%%%%%%%%
\section{Structural and elastic properties}\label{sec3}
%%%%%%%%%%%%%%%%%%%%%%%%%%%%%%%%%%%%%%%%%%%%%%%%%%%%%%%%%%%%%%%%%%
\subsection{Equilibrium structure}\label{sec3a}
%=================================================================

The equilibrium geometry was obtained by constant-volume relaxation and a subsequent fit of the resulting energy-over-volume curve to the Birch-Murnaghan equation of state (EOS)~\cite{Birch1947}. For the lattice parameters of unstrained hex-Ge, we found $a_0 = 3.996\Ang$, $c_0 = 6.592\Ang$, and the internal cell parameter $u_0 = 0.3743$. Both $u_0$ and the ratio of $c_0/a_0 = 1.6496$ are close to the values $(c/a)_\id=\sqrt{8/3}$ and $u_\id=3/8$ of the ideal lonsdaleite structure. The lattice parameters are in very good agreement with available experimental data~\cite{fadaly2020direct}. The EOS fit yields $B_0 = 67.6\GPa$ for the isothermal bulk modulus and $B_0' = 4.69$ for its pressure derivative.

%=====================================================================
\subsection{Symmetry-conserving strain} \label{sec3b}
%=====================================================================

For sufficiently small strains, the relation between the stress tensor $\pmb\sigma$ and the strain tensor $\pmb\epsilon$ is linear, i.e.\ Hooke's law
\begin{equation}\label{eq:hookeslaw}
\stress_i=\sum_{j=1}^6 \ElasticT_{ij}\, \strain_j
\end{equation}
holds, with the tensor of elastic constants $\ElasticT_{ij}$ in Voigt notation~\cite{Voigt1966} ($1,\ldots,6=xx,yy,zz,yz,zx,xy$). For hexagonal crystals, the symmetric elastic tensor has only five independent non-vanishing components~\cite{Nye1985, Wagner2002,Hanada2009}: $C_{11} = C_{22}$, $C_{33}$, $C_{12}$, $C_{13} = C_{23}$, $C_{44}=C_{55}$, and $C_{66} = \frac{1}{2}(C_{11}-C_{12})$.

First, we consider only normal stresses that leave the space-group symmetry unchanged. In this case, Hooke's law reduces to
\begin{subequations}\label{eq:hookeslaw_hex}
\begin{alignat}{1}
&\sigma_{xx}=\sigma_{yy}=(C_{11}+C_{12})\,\epsilon_{xx}+C_{13}\,\epsilon_{zz} \\
&\sigma_{zz}=2C_{13}\,\epsilon_{xx}+C_{33}\,\epsilon_{zz}
\end{alignat}
\end{subequations}
in hexagonal crystals, with the normal strains
\begin{subequations}
\begin{alignat}{1}
&\epsilon_{xx}=\epsilon_{yy}=(a-a_0)/a_0 \\
&\epsilon_{zz}=(c-c_0)/c_0
\end{alignat}
\end{subequations}
that are given by the deviations of the strained lattice constants $a$ and $c$ from their equilibrium values $a_0$ and $c_0$.

\begin{table*}
\centering
\caption{
Lattice parameters $a_0$ (in $\Ang$), $c_0$ (in $\Ang$), and $u_0$, elastic constants $C_{ij}$ (in $\GPa$), Young modulus $E$, biaxial modulus $Y$, isothermal bulk modulus $B_0$ (in $\GPa$) and its pressure derivative $B_0'$ as well as hydrostatic ratio $\RH$, biaxial ratio $\RB$, and Poisson ratio $\nu$ of hex-Ge. As explained in the text, the elastic properties have been obtained in two ways: directly calculated from the strained system and using \textsc{Elastic}.
}
\begin{ruledtabular}
\begin{tabular}{llllllllllllllllll}
                             & $a_0$    & $c_0$ & $u_0$ & $C_{11}{+}C_{12}$ & $C_{11}$ & $C_{12}$ & $C_{13}$ & $C_{33}$ & $C_{44}$ & $C_{66}$ & $E$  & $Y$ & $B_{0}$ & $B_0'$ & $\RH$ & $\RB$ & $\nu$ \\
\hline
Strain (PBEsol)              & $3.996$ & $6.592$ & $0.3743$ & $177.8$ &         &         & $23.4$  & $159.5$ &        &        & $153.3$ & $170.9$ & $67.6$ & $4.69$ & $1.023$ & $0.294$ & $0.132$ \\
\textsc{Elastic} (PBEsol)    &          &          &          & $177.7$ & $124.0$ & $53.7$  & $22.8$  & $159.4$ & $39.1$ & $35.2$ & $153.5$ & $171.2$ & $67.3$ &        & $0.967$ & $0.286$ & $0.128$ \\
\citeref{Fan2018}            & $4.030$  & $6.649$  &          & $156$   & $106$   & $50$    & $19$    & $150$   & $35$   & $28$   & $90$    &         & $60$   &        &         &         & $0.250$ \\
\citeref{Wang2003} (LDA)     &          &          &          & $193.1$ & $155.6$ & $37.5$  & $27.7$  & $169.3$ & $41.1$ & $59.1$ &         &         & $74.0$ &        &         &         &         \\
\citeref{fadaly2020direct} (exp.) & $3.9855$   & $6.5772$   &          &         &         &         &         &         &        &        &         &         &        &        &         &         &         \\
\end{tabular}
\end{ruledtabular}
\label{tab:elconst}
\end{table*}

\subsubsection{Hydrostatic pressure}

When a hydrostatic pressure $p$ is exerted on the system, the stress tensor reduces to $\stress_{ij} = -p\,\delta_{ij}$. Inserting this constraint into \equref{eq:hookeslaw_hex} yields the relation ${\strain_{zz}=\RH \strain_{xx}}$ between out-of-plane and in-plane strain, where $\RH$ is the hydrostatic ratio
\begin{equation}
	\label{eq:RH}
   \RH = \frac{(C_{11}+C_{12}) - 2C_{13}}{C_{33}-C_{13}}.
\end{equation}

The isothermal bulk modulus $B_0=-\left.V_0\frac{\mathrm{d}p}{\mathrm{d}V}\right|_{p=0}$ relates hydrostatic pressure to infinitesimal volume changes. For sufficiently small volume changes and pressures,
\begin{equation}\label{eq:hydrostrain}
p = -B_0\,\frac{\Delta V}{V_0} = -B_0\,(2\strain_{xx} + \strain_{zz})
\end{equation}
holds. Combining Eqs.~\eqref{eq:hookeslaw_hex}, \eqref{eq:RH}, and \eqref{eq:hydrostrain}, the bulk modulus is given by
\begin{equation}\label{eq:isothermalbulk}
   B_0 = \frac{(C_{11}+C_{12})C_{33} - 2(C_{13})^2}{(C_{11}+C_{12}) + 2(C_{33}-2C_{13})}.
\end{equation}

We simulate hex-Ge under hydrostatic pressure by a constrained relaxation of the atomic coordinates at a given fixed cell volume.

\subsubsection{Biaxial strain}

Biaxial strain perpendicular to the hexagonal $c$~axis is characterized by $\stress_{xx}=\stress_{yy}$ and $\stress_{zz}=0$, i.e.\ the forces along the $c$~axis vanish. Using \equref{eq:hookeslaw_hex}, the out-of-plane strain can be linked to the in-plane strain by $\strain_{zz}=-\RB\strain_{xx}$, where the biaxial ratio $\RB$ is given by
\begin{equation}\label{eq:biaxialratio}
   \RB = \frac{2C_{13}}{C_{33}}.
\end{equation}
The in-plane stress is related to the in-plane strain as $\stress_{xx} = Y \strain_{xx}$, with the biaxial modulus $Y$ that reads
\begin{equation} \label{eq:biaxialmodulus}
   Y = C_{11} + C_{12} - \frac{2C_{13}^2}{C_{33}}.
\end{equation}

To model biaxial strain perpendicular to the $c$~axis in hex-Ge, the lattice constant $a$ is fixed and all other degrees of freedom of the atomic geometry are determined by total-energy minimization. 

\subsubsection{Uniaxial strain}

For uniaxial strain along the $c$~axis, the in-plane normal stresses $\stress_{xx}=\stress_{yy}=0$ vanish and only $\stress_{zz}$ is nonzero. Inserting this condition into \equref{eq:hookeslaw_hex} yields the relation $\strain_{xx}=-\nu\strain_{zz}$ between the in-plane and out-of-plane strains with the Poisson ratio
\begin{equation}\label{eq:poisson1}
   \nu = \frac{C_{13}}{C_{11}+C_{12}}.
\end{equation}
The uniaxial stress $\stress_{zz} = E\strain_{zz}$ is related to the strain in this direction through the Young modulus $E$
\begin{equation}\label{eq:youngmodulus}
   E = C_{33} - \frac{2C_{13}^2}{C_{11}+C_{12}},
\end{equation}
which follows immediately from \equref{eq:hookeslaw_hex}.

Uniaxial strain can be modeled by fixing the $c$~lattice constant and relaxing all other atomic coordinates.

\subsubsection{Elastic constants}

For the situations of hydrostatic pressure, biaxial strain, and uniaxial strain, we determined the ratios between out-of-plane and in-plane strains for the smallest strains we investigated (about 1\,\%) and found $\RH = 1.023$, $\RB= 0.294$, and $\nu=0.132$, respectively.
Taking as additional input the bulk modulus $B_0=67.6\GPa$ from the EOS fit, the three elastic constants $C_{11}+C_{12}=177.8\GPa$, $C_{13}=23.4\GPa$, and $C_{33}=159.5\GPa$, the biaxial modulus $Y=170.9\GPa$, and the Young modulus $E=153.3\GPa$ can be computed.

%=====================================================================
\subsection{Symmetry-reducing strain}\label{sec3c}
%=====================================================================

All elastic constants $C_{ij}$ can be obtained by calculating normal and tangential stresses for specific deformations of the crystal lattice and solving Hooke's law, as implemented in the code \textsc{Elastic} \cite{elastic_code,Jochym1999}. The elastic constants obtained with \textsc{Elastic} confirm the values calculated for hex-Ge under hydrostatic pressure, biaxial strain, and unixaxial strain. A summary of structural parameters and elastic constants, including a comparison with results in literature, can be found in \tabref{tab:elconst}.

%%%%%%%%%%%%%%%%%%%%%%%%%%%%%%%%%%%%%%%%%%%%%%%%%%%%%%%%%%%%%%%%%%
\section{Electronic properties} \label{sec4}

%=====================================================================
\subsection{Symmetry-conserving strain} \label{sec4a}
%=====================================================================

\begin{figure*}
	\centering
	%\tikzsetnextfilename{Images_pdf/strain_bandstructures}
	%\input{Images/strain_bandstructures.tex}
    %\includegraphics[width=\textwidth]{Images_pdf/strain_bandstructures.pdf}
    \includegraphics[width=0.45\textwidth]{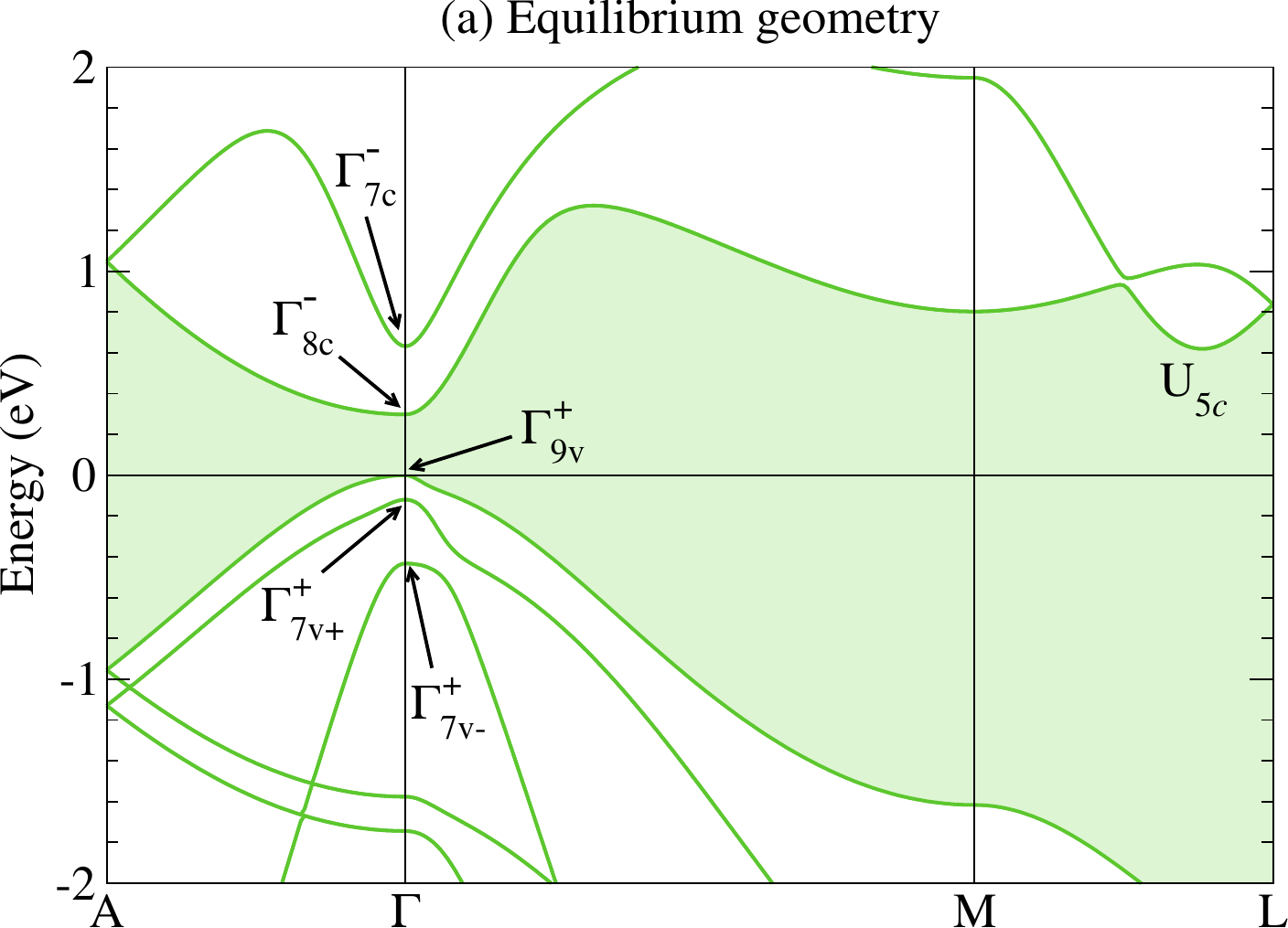} \hspace*{0.025\textwidth}
    \includegraphics[width=0.45\textwidth]{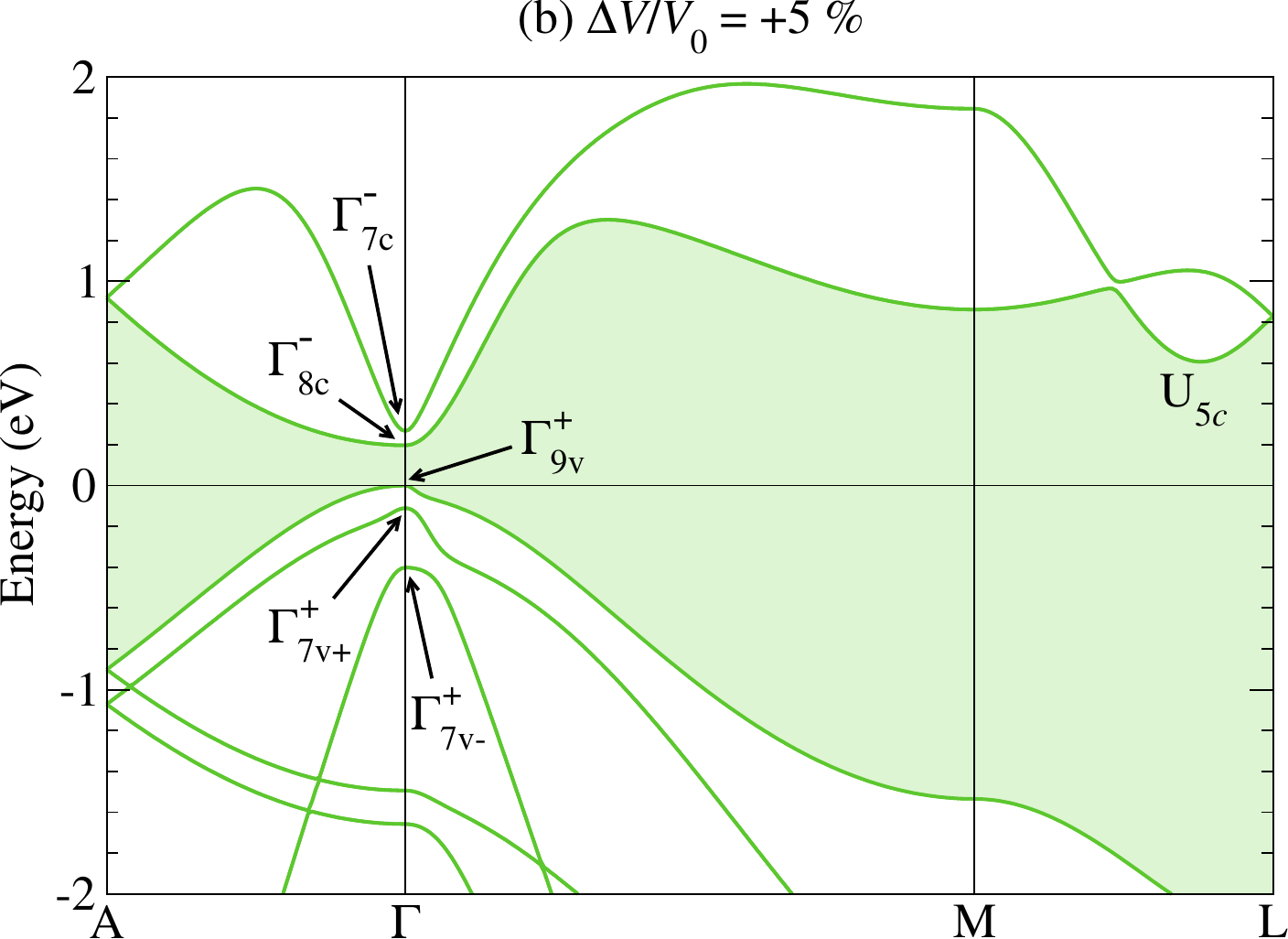} \\[\baselineskip]
    \includegraphics[width=0.45\textwidth]{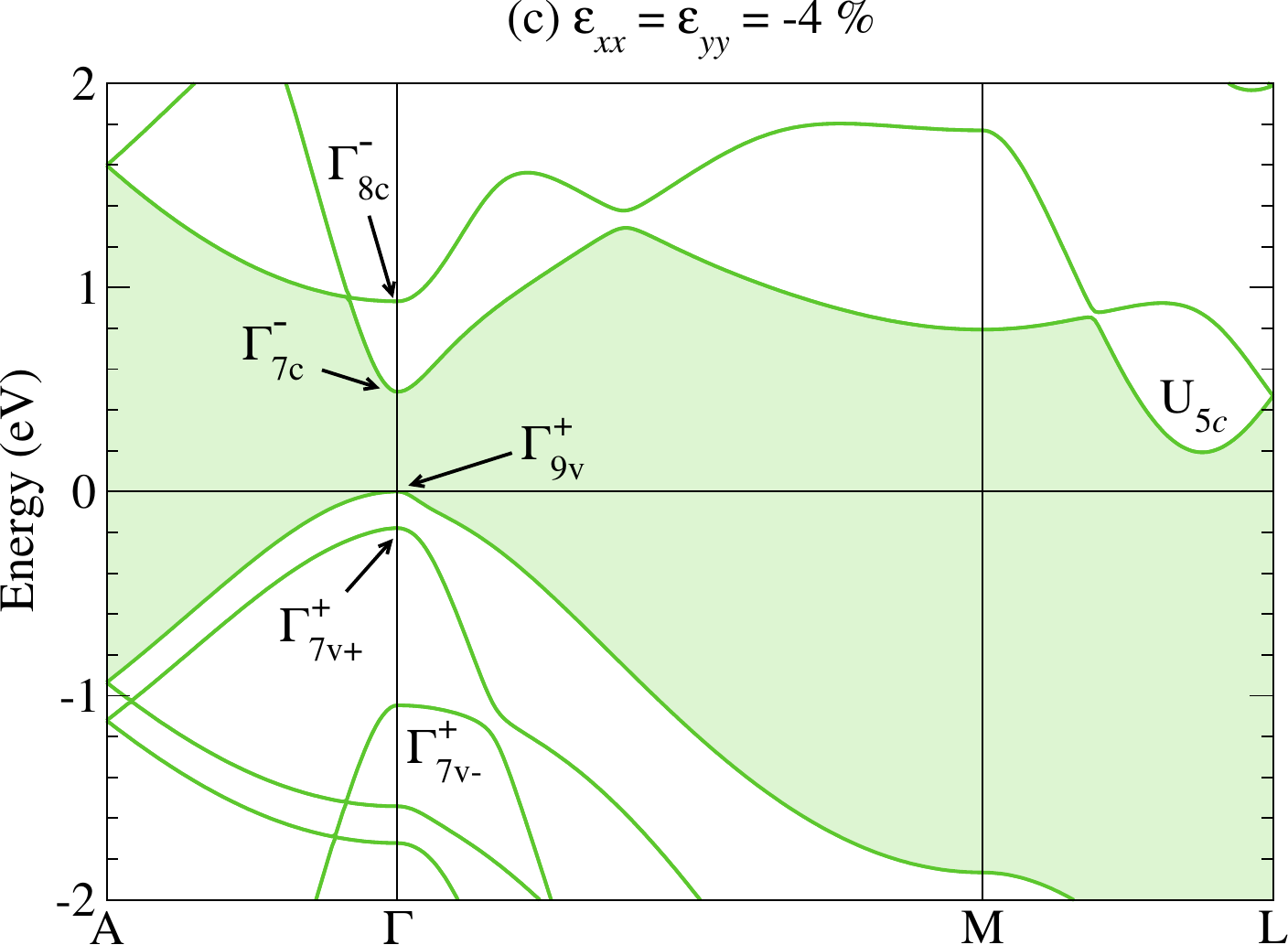} \hspace{0.025\textwidth}
    \includegraphics[width=0.45\textwidth]{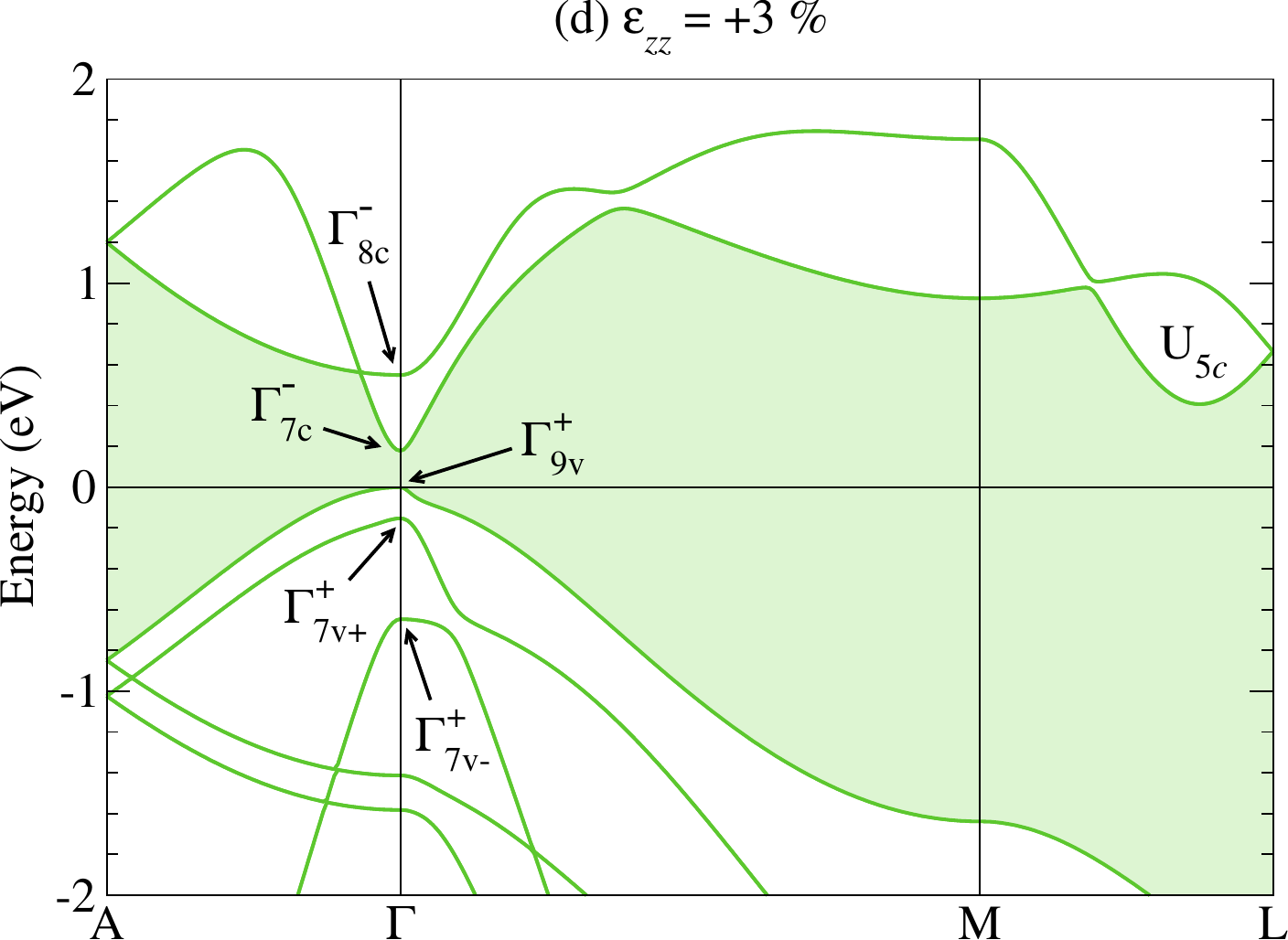}
	\caption{
		Band structure of hex-Ge for the unstrained equilibrium geometry (a), a volume change of $\Delta V / V_0 = +5\,\%$ (b), a biaxial strain of $\strain_{xx} = \strain_{yy} = -4\,\%$ (c), and a uniaxial strain of $\strain_{zz} = +3\,\%$ (d). The VBM is used as energy zero. The gap region is shaded.
	}
	\label{fig:bandstructure}
\end{figure*}

\begin{figure}
    \centering
    %\tikzsetnextfilename{Images_pdf/strain_evo}
    \includegraphics[width=\linewidth]{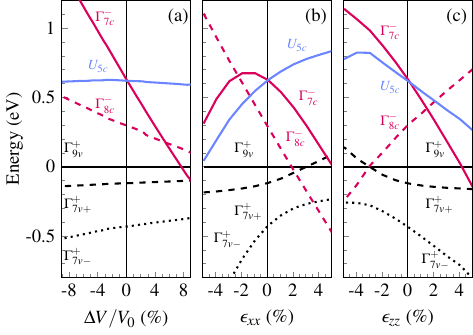}
    \caption{
        Variation of important electronic states in the vicinity of the band gap with hydrostatic pressure (a), biaxial strain (b), and uniaxial strain (c). The $\Gamma_{9v}^+$ level is used as energy zero. The three VBMs $\Gamma_{9v}^+$, $\Gamma_{7v+}^+$, and $\Gamma_{7v-}^+$ (black lines), the two CBMs $\Gamma_{8c}^-$ and $\Gamma_{7c}^-$ (red lines) as well as the CBM  $U_{5c}$ (blue line) are displayed.
    }
    \label{fig:bandstrain}
\end{figure}

The band structure of unstrained hex-Ge is displayed in \figref{fig:bandstructure}(a) along with the labels of relevant high-symmetry states~\cite{Roedl2019}. Besides the energy levels at the $\Gamma$~point, the $U_{5c}$ CBM is of particular interest, as it is close in energy to the CBMs at $\Gamma$. Under some strains, the $U_{5c}$ state on the $L$-$M$~line becomes the lowest conduction level and turns hex-Ge into an indirect semiconductor which is, of course, not desirable for high light-emission efficiency, as this CBM would act as a carrier trap. The strain dependence of the band-edge energies is plotted in \figref{fig:bandstrain} for hydrostatic pressure, biaxial strain, and uniaxial strain. Here, the $\Gamma_{9v}^+$ state, that is the valence-band maximum (VBM) for the unstrained structure, serves as reference level for all strain-induced energy shifts. 

\subsubsection{Hydrostatic pressure}

In general, we observe an increased splitting of all states at the $\Gamma$~point when hydrostatic pressure is exerted. Figure~\ref{fig:bandstrain}(a) shows that the VBMs at $\Gamma$ and the $U_{5c}$ CBM shift only marginally in the investigated range of volume changes. However, the energy of the two CBMs at $\Gamma$ increases strongly with rising pressure. The effect is particularly pronounced for the $\Gamma_{7c}^-$ state, as it forms an $sp$~gap with the $\Gamma_{9v}^+$ VBM~\cite{Raffy2002}.

Negative hydrostatic pressure corresponding to a volume increase of about $5\,\%$ would lead to the desired conduction-band inversion of the $\Gamma^-_{8c}$ and $\Gamma^-_{7c}$ states with a pseudodirect-to-direct gap transition at a gap of $0.17\eV$. In \figref{fig:bandstructure}(b), the band structure of hex-Ge for a volume dilatation of $5\,\%$ is shown, illustrating how the two CBMs approach. At even larger negative pressure, for $7.6\,\%$ volume increase, a semiconductor-to-metal phase transition occurs. Experimentally, negative hydrostatic pressures are hard, if not impossible, to realize, which is why this does not represent a promising route to engineer the electronic structure of hex-Ge. More sophisticated strains have to be explored.

\subsubsection{Biaxial strain}

Figure~\ref{fig:bandstrain}(b) shows that the band ordering in hex-Ge is more sensitive to biaxial strain perpendicular to the hexagonal $c$~axis than to hydrostatic pressure. The $\Gamma_{7c}^-$ state exhibits a highly nonlinear behavior leading to a conduction-band inversion for compressive biaxial strains larger than $2.2\,\%$. Unfortunately, a direct-to-indirect-semiconductor transition with a $\Gamma^+_{9v}\to U_{5c}$ fundamental gap occurs already at smaller compressive strains. The resulting band structure is illustrated in \figref{fig:bandstructure}(c) for $4\,\%$ of compressive biaxial strain. 

For tensile biaxial strain, the system becomes metallic at a strain of about $2\,\%$. The band ordering resembles that of zincblende HgTe or $\alpha$-Sn \cite{Kuefner2015} as a negative gap appears. Above $3\,\%$ of tensile strain, the band ordering changes to $\Gamma^+_{7v+} > \Gamma^+_{9v} > \Gamma^-_{8c}$. % and a small gap opens at $\Gamma$ between the empty $\Gamma^+_{7v+}$ and the filled $\Gamma^+_{9v}$ state. 
%It is accompanied by a phase transition from trivial to topological insulator \cite{Fu2007}.

\subsubsection{Uniaxial strain}

Uniaxial strain along the hexagonal $c$~axis offers great potential to engineer the electronic states, as the two CBMs $\Gamma_{8c}^-$ and $\Gamma_{7c}^-$ shift in opposite directions [see \figref{fig:bandstrain}(c)]. Already for a small tensile uniaxial strains of about $1.5\,\%$, a pseudodirect-to-direct gap transition occurs: hex-Ge becomes a direct-gap semiconductor with a gap of $0.4\eV$ and strong dipole transitions at the gap energy. The gap stays direct, but shrinks for larger strains until a semiconductor-to-metal transition occurs at about $4\,\%$ tensile strain. Similar behavior has been observed experimentally for wz-GaAs \cite{Signorello2014} and wz-GaP~\cite{Greil2016} under uniaxial strain. The strain range between $1.5$ and $4\,\%$ is likely to be experimentally accessible and further offers the possibility to tune the band gap between 0 and $0.4\eV$. Figure~\ref{fig:bandstructure}(d) shows the band structure of hex-Ge for $3\,\%$ of tensile uniaxial strain illustrating the direct band gap and the band inversion at $\Gamma$.

For tensile uniaxial strains above $4\,\%$ and compressive uniaxial strains above $3\,\%$, we predict a band inversion of conduction and valence bands.  

\subsubsection{Deformation potentials}

\begin{table}
\centering
\caption{Energies $E_0$ of high-symmetry states in the vicinity of the fundamental gap and the corresponding deformation potentials $\Xih$, $\Xib$, and $\Xiu$ for hydrostatic pressure, biaxial, and uniaxial strain, respectively. All quantities are given with respect to the VBM $\Gamma^+_{9v}$.}
\begin{ruledtabular}
\begin{tabular}{lrrrr}
State  & $E_0$ (eV) & $\Xih$ (eV) & $\Xib$ (eV) & $\Xiu$ (eV) \\
\hline
$\Gamma_{7v-}^+$ & $-0.433$ &  $0.51$  &  $8.93$   &  $-6.51$  \\
$\Gamma_{7v+}^+$ & $-0.120$ &  $0.15$  &  $2.73$   &  $-1.91$   \\
$\Gamma_{9v}^+$  & $0.000$  &  $0.00$  &  $0.00$   &  $0.00$    \\
$\Gamma_{8c}^-$  & $0.298$  &  $-1.95$ &  $-15.76$ &  $8.79$   \\
$\Gamma_{7c}^-$  & $0.633$  &  $-8.75$ &  $-6.83$  &  $-13.62$ \\
$U_{5c}$       & $0.620$  &  $-0.35$ &  $6.98$   &  $-6.93$  \\
\end{tabular}
\end{ruledtabular}
\label{tab:defpot}
\end{table}

In \tabref{tab:defpot}, the deformation potentials of the most important energy levels are compiled for the three considered symmetry-conserving strains. The deformation potentials $\Xi_j$ are defined as the linear expansion coefficients of the energy levels as a function of strain,
\begin{equation}
   E(\strain_j) \approx E_0 + \Xi_j\,\strain_j ,
\end{equation}
with $j=\text{h}$ for hydrostatic pressure ($\strain_\text{h}=\Delta V/V_0$), $j=\text{b}$ for biaxial strain ($\strain_\text{b}=\strain_{xx}=\strain_{yy}$), and $j=\text{u}$ for uniaxial strain ($\strain_\text{u}=\strain_{zz}$). Note in particular that for biaxial and uniaxial strain, the strain dependence of the energy levels quickly becomes nonlinear. In these cases, the explicitly calculated values for the larger strains can be taken from \figref{fig:bandstrain}. 

The volume deformation potentials are very small for the states $U_{5c}$, $\Gamma_{7v+}^+$, and $\Gamma_{7v-}^+$. The potential $\Xih = -8.7\eV$  of the $sp$~gap $\Gamma_{9v}^+\to\Gamma_{7c}^-$ is in excellent agreement with the value of $-8.8\eV$ for cubic Ge~\cite{Blacha1984}. 
It is also close to the value of $-8.25\eV$ computed for wz-GaAs~\cite{Cheiwchanchamnangij2011}. The volume deformation potential $\Xih = -1.9\eV$ for the $\Gamma_{9v}^+\to\Gamma_{8c}^-$ gap is significantly smaller. Also in this respect, hex-Ge behaves very similarly to wz-GaAs~\cite{Cheiwchanchamnangij2011}.

On average, the absolute values of the biaxial and uniaxial deformation potentials are larger than the volume deformation potentials, even when taking the relation $\Delta V/V_0\approx2\strain_{xx}+\strain_{zz}$ for the strain amplitudes into account. For all studied states except $\Gamma_{7c}^-$, the biaxial and uniaxial deformation potentials have opposite sign. However, their absolute values differ, as tensile (compressive) biaxial strain and compressive (tensile) uniaxial strain do not represent the same physical situation.

%=====================================================================
\subsection{Symmetry-reducing strain} \label{sec4b}
%=====================================================================

\begin{figure}
    \centering
    %\tikzsetnextfilename{Images_pdf/crystal_lattice}
    \includegraphics[width=.45\textwidth]{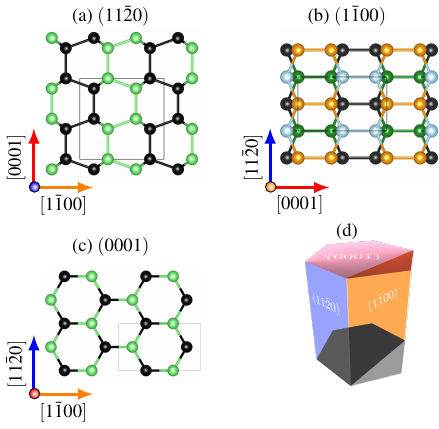}
    \caption{
        View of the orthorhombic supercell from three perspectives: side view on the $(11\bar{2}0)$ plane (a), side view on the $(1\bar{1}00)$ plane (b), and top view on the $(0001)$ plane (c). The orthorhombic supercell is indicated by black solid lines. Balls of different color symbolize Ge atoms belonging to different layers. The images of the atomic structures were generated with \textsc{Vesta}~\cite{vesta}. The orientation of the considered lattice planes is visualized in panel (d).
     \label{fig:unitcell}}
\end{figure}

\begin{table}
  \centering
  \caption{
    Lattice mismatch for pseudomorphic growth of hex-Ge on hex-Si and several III-V wurtzite semiconductors as possible substrates.
    The lattice mismatch is given as strain $\strain_{\perp} = (a_0^\mathrm{III-V} - a_0^\mathrm{Ge}) / a_0^\mathrm{Ge}$ and $\strain_{zz} = (c_0^\mathrm{III-V} - c_0^\mathrm{Ge}) / c_0^\mathrm{Ge}$.
    All substrate lattice parameters have been obtained from DFT with the PBEsol exchange-correlation functional, consistently with what has been done for hex-Ge.}

  \begin{ruledtabular}
  \begin{tabular}{lrrrr}
    Substrate & $a$ (\AA) & $c$ (\AA) & $\strain_{\perp}$ ($\%$) & $\strain_{zz}$ ($\%$) \\
    \hline
%    hex-Ge  & $3.996$ & $6.592$ &  $0.00$ &  $0.00$ \\
    hex-Si  & $3.826$ & $6.327$ & $-4.25$ & $-4.02$ \\
    wz-AlP  & $3.862$ & $6.341$ & $-3.35$ & $-3.81$ \\
    wz-AlAs & $4.003$ & $6.581$ & $+0.18$ & $-0.17$ \\
    wz-GaP  & $3.832$ & $6.317$ & $-4.10$ & $-4.17$ \\
    wz-GaAs & $3.988$ & $6.575$ & $-0.20$ & $-0.26$ \\
    wz-InP  & $4.148$ & $6.810$ & $+3.80$ & $+3.31$ \\
  \end{tabular}
  \end{ruledtabular}
  \label{tab:substrate}
\end{table}

\begin{figure*}
    \centering
    %\tikzsetnextfilename{Images_pdf/complex_map}
    \includegraphics[width=0.9\textwidth]{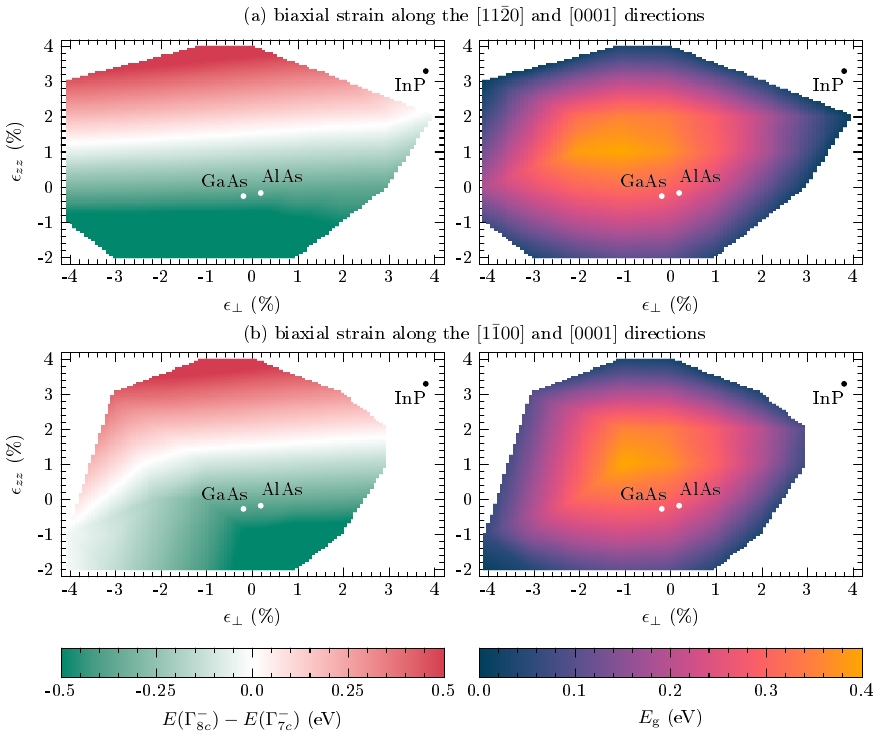}
    \caption{Conduction-band ordering at $\Gamma$ and size of fundamental band gap for biaxial strain in the $\{1\bar{1}00\}$ plane along the $[11\bar{2}0]$ and $[0001]$ directions (a) and in the $\{11\bar{2}0\}$ plane along the $[1\bar{1}00]$ and $[0001]$ directions (b). In the external white areas, where no data points are shown, the system is either metallic or the fundamental gap is indirect. The relative lattice mismatch of hex-Ge assuming pseudomorphic growth on wz-GaAs, wz-AlAs, and wz-InP substrates is indicated (see \tabref{tab:substrate}).}
    \label{fig:nonsymmband}
\end{figure*}

Pseudomorphic growth of hex-Ge on top of, e.g.\ wz-GaAs or wz-GaP substrates along the hexagonal $c$~axis does not permit to reach the preferential situation of tensile uniaxial strain. Growing instead on wz-GaAs or wz-GaP nanowire facets~\cite{Ikaros2017} represents a viable alternative to strain Ge along its hexagonal axis. However, the lattice mismatch between Ge and the nanowire substrate leads to a biaxial strain along the hexagonal $c$~axis and one of the $a$ or $b$~axes of the orthorhombic supercell shown in \figref{fig:unitcell}. The corresponding planes in the hexagonal cell are illustrated in \figref{fig:unitcell}, where also the top facet $\{0001\}$, i.e.\ the wire cross section, is indicated for comparison. Straining hex-Ge along the orthorhombic $a$-$c$ ($\{11\bar{2}0\}$ plane) or $b$-$c$ ($\{1\bar{1}00\}$ plane) facets reduces the space-group symmetry of hex-Ge which is why we refer to these strains as symmetry-reducing biaxial strains. For clarity, we use the same state labels as for the unstrained cells and the strained cells which conserve the full space group symmetry of the lonsdaleite structure.

The energy difference between the two lowest conduction bands, $\Gamma^-_{8c}$ and $\Gamma^-_{7c}$, as well as the size of the band gap at the $\Gamma$ point for symmetry-reducing biaxial strain are shown in \figref{fig:nonsymmband}. In view of laser applications, only strain configurations with a direct band gap are relevant, therefore only those will be discussed here. 

The conduction-band ordering for biaxial strain along the $\{1\bar{1}00\}$ facet is largely dominated by the strain component $\epsilon_{zz}$. The strain along the $[11\bar{2}0]$ direction has almost no impact. Consequently, the general picture is very similar to uniaxial strain. A tensile strain of $1\%$ -- $ 2\%$ yields the desired conduction-band inversion. Further increasing the strain reduces the value of the band gap until the material becomes metallic. 

For biaxial strain in the $\{11\bar{2}0\}$ plane, the conduction-band ordering is also dominated by the strain in $c$~direction. The conduction-band inversion occurs for $\epsilon_{zz}>1\%$ -- $2\%$. Also here, the band gap decreases with increasing strain. Only for compressive strains $\epsilon_\perp>2\,\%$ in the plane perpendicular to the $c$~axis, the $\epsilon_\perp$ component has an equally strong influence and the conduction-band inversion occurs for lower $\epsilon_{zz}$. However, in this region, the gap is already very small. Therefore, we can conclude that the desired band ordering, which can be obtained for uniaxial strain, is stable with regard to small additional strains perpendicular to the $c$~axis.

We also checked whether a particular substrate can lead to lattice strain in the interesting regime. Some wurtzite substrates and their lattice mismatch relative to hex-Ge are listed in \tabref{tab:substrate}, and the resulting strain of the most interesting substrates is  indicated in \figref{fig:nonsymmband}. None of the considered substrates could induce the desired results, but one could use either wz-GaAs or wz-AlAs as a substrate to grow almost strain-free hex-Ge. Wz-InP (wz-GaP, hex-Si) can serve as substrate for tensile (compressive) strain in hex-Ge overlayers.

%%%%%%%%%%%%%%%%%%%%%%%%%%%%%%%%%%%%%%%%%%
\section{Optical properties} \label{sec5}
%%%%%%%%%%%%%%%%%%%%%%%%%%%%%%%%%%%%%%%%%%

Always in view of optoelectronic applications, we analyze how the pseudodirect-to-direct band-gap transition induced by tensile uniaxial strain affects the optical properties near the fundamental band edge.

We calculate the dielectric tensor components $\varepsilon_{ii}(\omega)$ in the independent-particle approximation using the optical matrix elements $\langle c\mathbf{k}|\mathbf{p}|v\mathbf{k}\rangle$ of the momentum operator $\mathbf{p}$ between conduction band $c$ and valence band $v$ at a given $\mathbf{k}$-point applying the longitudinal gauge \cite{Gajdos2006}. The diagonal elements of the imaginary part of the dielectric tensor are given by
\begin{equation}
\begin{split}
  \IM\varepsilon_{ii}(\omega) = \frac{1}{\Omega_0}\frac{\pi e^2}{\epsilon_0 m^2\hbar}
                                  \sum_{cv\mathbf{k}}& w_{\mathbf{k}} \frac{|\langle c\mathbf{k}| p_i | v\mathbf{k}\rangle |^2}{(\omega_{c\mathbf{k}} - \omega_{v\mathbf{k}})^2 } \times \\
                                  &\delta(\omega_{c\mathbf{k}} - \omega_{v\mathbf{k}} - \omega),
\end{split}
\end{equation}
with the valence and conduction band energies $\hbar \omega_{v\mathbf{k}}$ and $\hbar \omega_{c\mathbf{k}}$. The unit-cell volume is denoted by $\Omega_0$ and $w_\mathbf{k}$ is the $\mathbf{k}$-point weight. In the calculations, the $\delta$~functions are represented by Lorentzians with a width of 0.05~eV. Excitonic effects can be neglected here. The dielectric constant of Ge is large, resulting in strong screening of the electron-hole interaction. Consequently, the exciton binding energy of cubic Ge amounts to only few meV \cite{altarelli_lipari_1976} and similar values can be expected for the hexagonal structure. What is more, the symmetries of the band-edge states, which crucially govern the low-energy absorption and emission properties of hexagonal Ge, are not affected by the inclusion of the electron-hole interaction.

%We start the analysis of the results considering the optical transition matrix elements at $\Gamma$ as a function of uniaxial strain along the $c$~axis (see \figref{fig:omeuniax}). We can see that light-induced transitions from the three highest valence bands into the $\Gamma^-_{7c}$ conduction band are significantly stronger than transitions into $\Gamma^-_{8c}$. 

%For light polarized perpendicular to the $c$~axis, the optical matrix elements of transitions into $\Gamma^-_{7c}$ are about two orders of magnitude stronger than those into $\Gamma^-_{8c}$. The strength of transitions into the $\Gamma^-_{7c}$ band slowly decreases with tensile uniaxial strain, whereas the matrix elements of the transitions into the $\Gamma^-_{8c}$ band behave differently with strain. For light polarized parallel to the $c$~axis,  $\Gamma^+_{7v+}\rightarrow\Gamma^-_{7c}$ and $\Gamma^+_{7v-}\rightarrow\Gamma^-_{7c}$ are the dominating transitions over the whole range of investigated uniaxial strain. Going from compressive to tensile uniaxial strain, the comparably small $\Gamma^+_{9v}\rightarrow\Gamma^-_{7c}$ matrix element and the $\Gamma^+_{7v+}\rightarrow\Gamma^-_{7c}$ the matrix element decrease, whereas the matrix element of the $\Gamma^+_{7v-}\rightarrow\Gamma^-_{7c}$ transition remains almost constant. The transition matrix elements to the $\Gamma^-_{8c}$ band are over four orders of magnitudes smaller.

\begin{figure*}
    \centering
    %\tikzsetnextfilename{Images_pdf/ImEps2}
    \includegraphics[width=\linewidth]{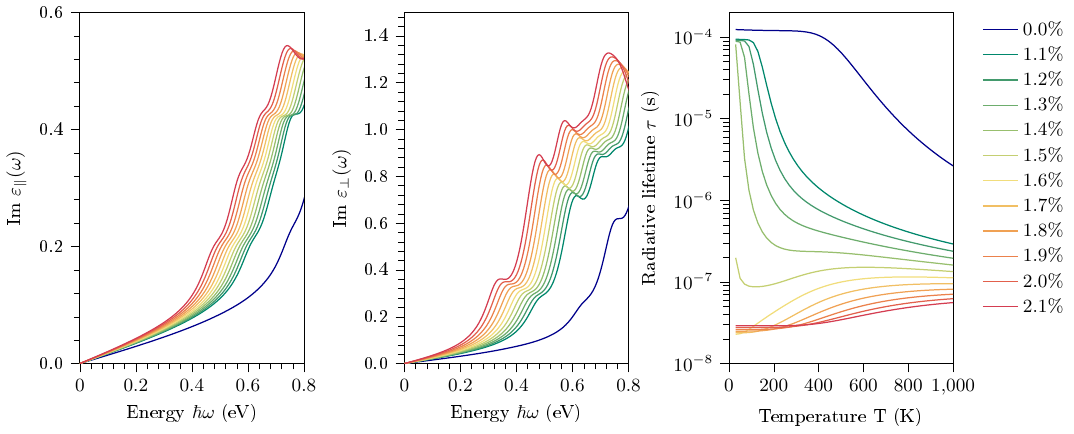}
    \caption{
      Imaginary part of dielectric function and radiative lifetime for increasing uniaxial strain along the hexagonal $c$~axis. The blue curve represents the reference of unstrained hex-Ge, whereas the curves ranging from green to red indicate increasing uniaxial strain from $1.1\,\%$ up to $2.1\,\%$.}
    \label{fig:imepsuniax}
\end{figure*}

The global effect of tensile uniaxial strain on optical properties is illustrated in \figref{fig:imepsuniax}, where we show the imaginary part of the frequency-dependent dielectric tensor components $\epsilon_{jj}(\omega)$ ($j={}\perp,\parallel$) and the temperature-dependent radiative lifetimes $\tau$ which are calculated as detailed in Ref.~\cite{Roedl2019}. 

Unstrained hex-Ge does not exhibit an absorption peak at the fundamental band gap of $0.3\,\eV$. The first peak rather occurs at $0.6\,\eV$, the energy of the strong dipole transition between the highest valence band and the $\Gamma_{7c}^-$ state. For increasing tensile uniaxial strain, this absorption peak shifts to lower energies, until it coincides with the band-gap energy for tensile uniaxial strains above $1.5\,\%$ as a consequence of the conduction-band inversion. For strains beyond the band-inversion point, the slope of the absorption edge is much steeper which is typical for a direct semiconductor with dipole-allowed transitions at the absorption edge.
The analysis of the optical transition matrix elements at $\Gamma$ as a function of uniaxial strain along the $c$-axis reveals an overall rather weak effect on the optical matrix elements. The smaller band gap for increasing strain induces a larger value of the imaginary dielectric function at the absorption edge, that is not compensated by the concomitant slight decrease of the optical matrix element of the $\Gamma^+_{9v}\rightarrow\Gamma^-_{7c}$ transition.

The occurrence of band inversion for moderate uniaxial strain along the $c$-axis is also obvious from the behaviour of the radiative lifetime. Tensile uniaxial strain above $1.5\,\%$ gives rise to drastic changes of the radiative lifetime. Below $1.5\,\%$ of strain, the lifetime curve shows the typical behavior of a system with two decay channels: a low-energy transition with small matrix element and a high-energy transition with large matrix element \cite{Roedl2019}. Upon increasing temperature, the impact of the high-energy transition becomes stronger and the lifetime drops by several orders of magnitude. The large low-temperature lifetime of $\sim10
^{-4}\,\text{s}$ is typical for a non-emitting system, the high-temperature value of $\sim10
^{-7}\,\text{s}$ rather corresponds to a direct semiconductor. For uniaxial strains above the pseudodirect-to-direct gap transition, the lifetime is of the order of $\sim10^{-8}$ -- $10^{-7}$\,s and largely constant over the entire temperature range. This behaviour is characteristic for a system with one dominating decay channel as the band-edge transition in direct semiconductors~\cite{Roedl2019} and shows that the strong dipole transition at the band gap outweighs all the rest.

The optical properties of uniaxially strained hex-Ge show that the material is an efficient light absorber with a dipole active direct-gap transition and low radiative lifetimes. Therefore, it is an interesting candidate material for silicon-technology based active optical devices which can compete with the light-emission efficiency of present-day III-V semiconductor devices.

%\begin{figure}
%    \centering
%    %\tikzsetnextfilename{Images_pdf/ome_uniax}
%    \includegraphics[width=\linewidth]{Images_pdf/ome_uniax.pdf}
%    %\input{Images/ome_uniax.tex}
%    \caption{
%      Optical matrix elements of transitions from the three highest valence bands to the two lowest conduction bands at $\Gamma$ as %function of uniaxial strain along the hexagonal $c$~axis. The vertical bar indicates the critical strain for which band inversion is %achieved. The top panel is for light polarized perpendicular, the bottom one for light polarized parallel to the hexagonal $c$~axis.}
%    \label{fig:omeuniax}
%\end{figure}

%%%%%%%%%%%%%%%%%%%%%%%%%%%%%%%%%%%%%%%%%%%%%%%%%%%%%%%%%%%%%%%%%%
\section{Summary and conclusions}\label{sec6}
%%%%%%%%%%%%%%%%%%%%%%%%%%%%%%%%%%%%%%%%%%%%%%%%%%%%%%%%%%%%%%%%%%

In summary, we have explored various possibilities to turn hexagonal Ge in the lonsdaleite phase with its pseudodirect band gap into a direct semiconductor with light-emission efficiencies suitable for technological applications. To this end, we investigated the impact of hydrostatic pressure, biaxial strain, and uniaxial strain on the electronic structure and the optical properties of hex-Ge. Hydrostatic pressure turns out unsuitable to obtain the desired modification of the conduction-band ordering and the concomitant pseudodirect-to-direct gap transition, as it occurs only for technically not accessible negative pressures. For biaxial strain within the hexagonal plane, either an insulator-to-metal transition occurs or the material becomes indirect before the conduction-band ordering changes.

For moderate tensile uniaxial strains between $1.5\,\%$ and $4\,\%$, the desirable band inversion between the $\Gamma^-_{8c}$ and $\Gamma^-_{7c}$ conduction states is predicted by our calculations. The resulting gaps vary between $0.4$ and $0\,\eV$. Such strains are within reach of experimental realization. We could also show that small additional strain within the hexagonal plane, as it may occur in practical technological growth processes, has only little impact on the electronic structure. Uniaxially strained hex-Ge could, for instance, be pseudomorphically grown on the facets of III-V wurtzite-semiconductor nanowires.

The inversion of the two lowest conduction minima at $\Gamma$ makes hexagonal Ge an excellent absorber or emitter, with a radiative lifetime smaller by three order of magnitudes than the one of the unstrained material. Provided that a suitable growth method can be found, hex-Ge can be used as a direct light emitter compatible with CMOS technology. Another unexplored route that deserves future attention is the combination of strain with alloying of Ge and Si to control the size of the band gap and the conduction-band ordering at $\Gamma$ at the same time. This route is particularly promising considering that light emission form hexagonal SiGe nanowires has recently been  demonstrated~\cite{fadaly2020direct} and that the control of shape and strain of crystalline nanowires keeps showing great progress.~\cite{Assali2019}

%%%%%%%%%%%%%%%%%%%%%%%%%%%%%%%%%%%%%%%%%%%%%%%%%%%%%%%%%%%%%%%%%%
\begin{acknowledgments}
This work is supported by the European Commission in the framework of the H2020 FET Open project SiLAS (Grant agreement No.~735008). C.\,R.\ acknowledges financial support from the Marie Sk\l{}odowska-Curie Actions (Grant agreement No.~751823). Computing time was granted by the Leibniz Supercomputing Centre on SuperMUC (project No.~pr62ja).
\end{acknowledgments}

\bibliography{references.bib}

% \appendix

\end{document}